\documentclass[aps,prl,preprint,groupedaddress,amsmath]{revtex4-2}
\usepackage{graphicx}
\usepackage{dcolumn}
\usepackage{bm}
\usepackage{hyperref}
\usepackage{xcolor}
\usepackage{ctable}

\begin{document}


\title{Critical temperature of the homogeneous repulsive weakly interacting Bose gas}


\author{Nguyen Van Thu}
\email[]{nvthu@live.com}
\affiliation{Department of Physics, Hanoi Pedagogical University 2, Hanoi, Vietnam}
\author{Pham Duy Thanh}
\affiliation{Department of Physics, Hanoi Pedagogical University 2, Hanoi, Vietnam}

\date{\today}

\begin{abstract}
Beyond standard approaches in existing literature, we explore the relative shift in the transition temperature of a homogeneous dilute Bose gas using the Cornwall–Jackiw–Tomboulis effective action formalism within the improved Hartree-Fock approximation. The first correction to the enhancement of the transition temperature, as compared to that of the ideal Bose gas, is expressed in a universal form: it is linear with respect to the scattering length. The slope of this linear relationship shows excellent agreement with the exact numerical calculations presented in previous studies. Additionally, we identify non-universal terms contributing to the shift.
\end{abstract}


\maketitle

\section{}
Since the first experimental observation of a Bose-Einstein condensate \cite{Anderson1995}, numerous studies have been conducted to achieve the condensation of Bose gases in laboratory settings \cite{Wilson2021,Burchianti2018,Lopes2017,Navon2021}. However, the theoretical calculations regarding the effect of interatomic interactions on the transition temperature of a homogeneous repulsive weakly interacting Bose gas have been the subject of a prolonged and controversial history \cite{Lee1958,Huang1999,Mihaila2011}. Numerous studies yield two common conclusions: (i) in the limit of the homogeneous weakly interacting Bose gas, the model with pairwise through the hard-sphere potential contact potential is appropriate; (ii) the relative shift in the transition temperature $T_C$ of the homogeneous repulsive weakly interacting Bose gas, as compared to that of the ideal Bose gas, in the first-order approximation of the $s$-wave scattering length $a_s$, can be expressed in a universal form
\begin{eqnarray}
\frac{\Delta T_C}{T_C}\equiv\frac{T_C-T_0}{T_C}=c.\rho^{1/3}a_s,\label{universal}
\end{eqnarray}
in which $\rho$ is the fixed particle density and $c$ is a constant. The critical temperature of the ideal Bose gas $T_0=\frac{2\pi\hbar^2}{2mk_B}\left[\frac{\rho}{\zeta(3/2)}\right]^{2/3}$ with $m, \rho$ being the atomic mass and particle density, respectively; $\zeta(x)$ is the zeta function. While authors of Refs. \cite{Toyoda1982,Wilkens2000} predicted nagative values of $c$ some other results showed positive values \cite{Bijlsma1996,Baym1999,Holzmann1999}. Most recent achievements have pointed out an increase of the critical temperature \cite{Arnold2001,Kashurnikov2001,Baym2001,Holzmann2004} owning to the change of the velocity profile just above the transition point and, consequently, the proportion of atoms with low velocities is higher than in the ideal gas \cite{Holzmann1999a}. Nevertheless, all subsequent calculations, alas, yield different values for the constant $c$ in a large range. To the best of our knowledge, the smallest reported positive value is $c=0.34\pm 0.06$, obtained by the authors of Ref. \cite{Grueter1997} using the path-integral Monte Carlo simulations and finite-size scaling. Conversely, utilizing the nonperturbative optimized linear $\delta$ expansion method applied to the $O(N)$ scalar field model, Cruz and co-workers \cite{SouzaCruz2001} showed $c=3.059$ for $N=2$. 

Beyond the available methods in literature, in this Letter we use the Cornwall-Jackiw-Tomboulis (CJT) effective action approach \cite{Cornwall1974} developed for the dilute Bose gas \cite{Thu2024} to investigate the effect of the interatomic interaction on the shift of the critical temperature. To this end, we employ the path-integral formalism, and the corresponding Euclidean partial function of the model system is ${\cal Z}=\int{\cal D}[\psi,\psi^*]\exp\left\{-S[\psi,\psi^*]/\hbar\right\}$ with the action functional $S[\psi,\psi^*]=\int dtd\vec{r}{\cal L}$  \cite{Pethick2008,Pitaevskii2003}.  Without any external field, The Lagrangian density is expressed in terms of the field operator $\psi$, chemical potential $\mu$ as below
\begin{equation}
{\cal L}=\psi^*\left(-i\hbar\frac{\partial}{\partial t}-\frac{\hbar^2}{2m}\nabla^2\right)\psi-\mu\left|\psi\right|^2+\frac{g}{2}\left|\psi\right|^4,\label{eq:1}
\end{equation}
where $g=4\pi\hbar^2a_s/m$ is the atomic mass and the coupling constant. We restrict our consideration for the case of the repulsive interatomic interaction. In order to consider in the Hartree-Fock approximation (HFA), the field operator is decomposed into its expectation value $\psi_0$ and two real fields $\psi_1$ and $\psi_2$ following the form $\psi\rightarrow \psi_0+(\psi_1+i\psi_2)/\sqrt{2}$. For the system under consideration, $\psi_0$ is real and plays the role of the order parameter whereas $\psi_1,\psi_2$ correspond to the fluctuations of the field. By this decomposition, Eq. (\ref{eq:1}) yields the interacting Lagrangian density in the two-loop approximation \cite{VanThu2022},
\begin{equation}
{\cal L}_{\rm int}=\frac{g}{2}\psi_0\psi_1(\psi_1^2+\psi_2^2)+\frac{g}{8}(\psi_1^2+\psi_2^2)^2.\label{eq:Lint}
\end{equation} 
The CJT effective potential in the HFA can be read off from (\ref{eq:Lint})
\begin{eqnarray}
V_\beta^{\rm{(CJT)}} =&&-\mu\psi_0^2 +\frac{g}{2}\psi_0^4+\frac{1}{2}\int_\beta \mbox{tr}\left[\ln G^{-1}(k)+D_0^{-1}(k)G(k)-{1\!\!1}\right]\nonumber\\
&&+\frac{3g}{8}(P_{11}^2+P_{22}^2)+\frac{g}{4}P_{11}P_{22},\label{eq:VHF}
\end{eqnarray}
in which $D_0(k)$ is the free propagator and $G(k)$ is the propagator in the HFA, they depend on the wave vector $\vec{k}$. The notation $\int_\beta$ encapsulates the integration over momentum space combined with a summation over bosonic Matsubara frequencies, specifically given by $\int_\beta=\frac{1}{\beta}\sum_{n=-\infty}^{+\infty} \int \frac{d^3\vec{k}}{(2\pi)^3}$; the Matsubara frequency for the boson $\omega_n=2\pi nk_BT\equiv \frac{2\pi n}{\beta}$. The functions
\begin{eqnarray}
P_{aa}\equiv \int_\beta G_{aa}(k),\label{integral}
\end{eqnarray}
are termed the momentum integrals.

It is widely recognized that the condensation of Bose gas is associated with the spontaneous breaking of the continuous symmetry $U(1)$ \cite{Anderson2018}. This leads to a gapless dispersion relation for excitations, as required by the Goldstone theorem \cite{Goldstone1961}. In the context of Bose-Einstein condensates, this theorem is referred to as the Hugenholtz–Pines theorem \cite{Hugenholtz1959} at zero temperature and the Hohenberg-Martin theorem \cite{Hohenberg1965} at finite temperature. However, the CJT effective potential (\ref{eq:VHF}) fails to produce the Goldstone boson. To address this issue, an additional phenomenological term must be incorporated into this CJT effective potential. This additional term has been shown to be unique for a specific system, and in our case, it modifies the CJT effective potential (\ref{eq:VHF}) to become \cite{VanThu2022},
\begin{eqnarray}
    \label{eq:VIHF}
\widetilde{V}_\beta^{\rm{(CJT)}} =&&  -\mu\psi_0^2 +\frac{g}{2}\psi_0^4+\frac{1}{2}\int_T \mbox{tr}\left[\ln D^{-1}(k)+D_0^{-1}(k)D(k)-{1\!\!1}\right]\nonumber\\
&&+\frac{g}{8}(P_{11}^2+P_{22}^2)+\frac{3g}{4}P_{11}P_{22}.
\end{eqnarray}
Minimizing this potential with  respect to $\psi_0$ one arrives at the gap equation
\begin{equation}
-\mu+g\psi_0^2+\frac{3g}{2}P_{11}+\frac{g}{2}P_{22}=0.\label{eq:gap}
\end{equation}
Similarly, the Schwinger-Dyson can be obtained
\begin{equation}
M=-\mu+3g\psi_0^2+\frac{g}{2}P_{11}+\frac{3g}{2}P_{22},\label{eq:SD}
\end{equation}
for the effective mass $M$. The new propagator in Eq. (\ref{eq:VIHF}) has the form
\begin{equation}
D(k)=\frac{1}{\omega_n^2+E^2(k)}\left(
              \begin{array}{lr}
                \frac{\hbar^2k^2}{2m} & \omega_n \\
                -\omega_n & \frac{\hbar^2k^2}{2m}+M \\
              \end{array}
            \right).\label{eq:proIHF}
\end{equation}
It is obvious that the Goldstone is restored with the gapless dispersion relation
\begin{equation}
E(k)=\sqrt{\frac{\hbar^2k^2}{2m}\left(\frac{\hbar^2k^2}{2m}+M\right)}.\label{disperIHF}
\end{equation}
Within the CJT effective potential and relating equations (\ref{eq:VIHF})-(\ref{disperIHF}), we are in a so-called improved Hartree-Fock approximation (IHFA). To maintain consistency within our theoretical framework, we impose the condition that the chemical potential is equal to the first derivative of the pressure with respect to the particle density. Here, the pressure is defined as the negative of the CJT effective potential (\ref{eq:VIHF}) evaluated at its minimum, thereby satisfying both the gap and SD equations. Therefore the chemical potential gives 
\begin{equation}
\mu=g\psi_0^2+gP_{11}.\label{eq:chemical}
\end{equation}

We now come back to our main aim in this Letter, which is to calculate the influence of the interparticle interaction on the transition temperature of the weakly interacting Bose gas. As we know, state of our system is governed by equations of state called the gap and SD equations (\ref{eq:gap}) and (\ref{eq:SD}) satisfying the consistent condition (\ref{eq:chemical}). To proceed further, we now compute the momentum integrals. In the IHFA, these integrals are evaluated using Eq. (\ref{integral}), with $G(k)$ replaced by $D(k)$. By substituting (\ref{eq:proIHF}) into (\ref{integral}) and applying the rule for summation over the Matsubara frequency \cite{Schmitt2010}, these momentum integrals are formulated
\begin{eqnarray}
P_{11}=&&\int\frac{d^3\vec{k}}{(2\pi)^3}\frac{\hbar^2k^2/2m}{2E(k)}+\int\frac{d^3\vec{k}}{(2\pi)^3}\frac{\hbar^2k^2/2m}{E(k)\left[e^{\beta E(k)}-1\right]},\nonumber\\
P_{22}=&&\int\frac{d^3\vec{k}}{(2\pi)^3}\frac{\hbar^2k^2/2m+M}{2E(k)}+\int\frac{d^3\vec{k}}{(2\pi)^3}\frac{\hbar^2k^2/2m+M}{E(k)\left[e^{\beta E(k)}-1\right]}.\label{tichphan2}
\end{eqnarray}
The first terms in right-hand side of (\ref{tichphan2}) are not explicitly dependent of temperature and ultraviolet divergent. By using the standard dimensional regularization \cite{Andersen2004}, this divergence can be removed. Just below the critical temperature, the effective mass approaches zero thus $\beta M\ll 1$. Accounting the leading term of the scattering length, the second terms in right-hand side of (\ref{tichphan2}) are computed by applying the Bose integral \cite{Olver2010}. At a finite temperature $T$, the condition for diluteness requires that the scattering length is not only much smaller than the interparticle distance but also considerably smaller than the de Broglie wavelength $\lambda_B=\sqrt{2\pi\hbar^2/mk_BT}$, i.e., $\rho^{1/3}a_s\ll 1$ and $\frac{a_s}{\lambda_B}\ll1$ \cite{Andersen2004,Shi1998}. This fact together with the equations of state (\ref{eq:gap}), (\ref{eq:SD}) and the consistent condition (\ref{eq:chemical}) give 
\begin{widetext}
\begin{eqnarray}
-1+\frac{\rho_0}{\rho}+\frac{2\sqrt{2}}{3\sqrt{\pi}}{\cal M}^{3/2}(\rho a_s^3)^{1/2}+\frac{\zeta(1/2)a_s{\cal M}}{\lambda_B}+\frac{\zeta(3/2)}{\rho\lambda_B^3}&=&0,\nonumber\\
-1+\frac{3\rho_0}{\rho}-\frac{10\sqrt{2}}{3\sqrt{\pi}}{\cal M}^{3/2}(\rho a_s^3)^{1/2}+\frac{5\zeta(1/2)a_s{\cal M}}{\lambda_B}+\frac{\zeta(3/2)}{\rho\lambda_B^3}&=&{\cal M},\label{eqnew1}
\end{eqnarray}
with $\rho_0\equiv \psi_0^2$ the condensate density and ${\cal M}=M/(g\rho)$ the dimensionless effective mass. The solution for Eq. (\ref{eqnew1}) is approximated
\begin{eqnarray}
\frac{\rho_0}{\rho}\approx  1-\frac{8}{3\sqrt{\pi}}(\rho a_s^3)^{1/2}-\frac{2\zeta(1/2)a_s}{\lambda_B}-\frac{\zeta(3/2)}{\rho\lambda_B^3}.\label{x}
\end{eqnarray}
\end{widetext}
It is very conspicuous to see a second order phase transition as it should be \cite{Andersen2004}. At the transition point the condensed density vanishes. The Eq. (\ref{x}) yields the critical temperature
\begin{widetext}
\begin{eqnarray}
T_C\approx  \frac{2\pi\hbar^2}{mk_B}\left[\frac{n_0}{\zeta(3/2)}\right]^{2/3}\left[1-\frac{4\zeta(1/2)}{3\zeta(3/2)^{1/3}}(\rho a_s^3)^{1/3}\right].\label{Tc}
\end{eqnarray}
\end{widetext}
Plugging (\ref{Tc}) into (\ref{universal}) one has
\begin{eqnarray}
\frac{\Delta T_C}{T_0}=-\frac{4\zeta(1/2)}{3\zeta(3/2)^{1/3}}\rho^{1/3}a_s.\label{shift}
\end{eqnarray}
It is obvious that $c=-\frac{4\zeta(1/2)}{3\zeta(3/2)^{1/3}}=\sqrt{2}\approx 1.414$ in our method. This result aligns closely with the value $c=1.32\pm 0.02$ as reported in Ref. \cite{Arnold2001}, which was derived from measuring the numerical coefficient in a lattice simulation of $O(2)$ scalar field theory. Similarly, good agreement is observed with the result $c=1.27\pm0.11$, as presented in Ref. \cite{Kastening2004}, where variational perturbation theory through seven loops was employed. Accounting for higher-order terms of the gas parameter, our result can be expressed as 
\begin{eqnarray}
\frac{\Delta T_C}{T_0}=-\frac{4\zeta(1/2)}{3\zeta(3/2)^{1/3}}\rho^{1/3}a_s+\frac{8}{9\sqrt{\pi}}\rho^{1/2}a_s^{3/2}+\frac{4\zeta(1/2)^2}{9\zeta(3/2)^{2/3}}\rho^{2/3}a_s^2+{\cal O}(\rho a_s^3).\label{high}
\end{eqnarray}
The second correction, which demonstrates that $\Delta T$ is proportional to $\sqrt{a_s^3}$, indicates that it cannot be extended to the region where $a_s<0$. This provides evidence that a gas with attractive interactions will undergo collapse, thereby invalidating the scattering-length description. The condition of diluteness suggests that the contribution from higher-order gas parameter terms has a minimal effect on the shift of the critical temperature. For comparison purposes, Eq. (\ref{high}) may be reformulated as follows
\begin{eqnarray}
\frac{\Delta T_C}{T_0}\approx&&1.414\rho^{1/3}a_s+0.501\rho^{1/2}a_s^{3/2}+0.499\rho^{2/3}a_s^2+{\cal O}(\rho a_s^3).\label{high1}
\end{eqnarray}
It is of interest to compare Eq. (18) with the result obtained in Refs. \cite{Holzmann2001,Arnold2001a}, where perturbative expansions and analytic techniques yield 
\begin{eqnarray}
\frac{\Delta T_C}{T_0}\approx&& 1.32\rho^{1/3}a_s+19.7518\rho^{2/3}a_s^2\ln(\rho^{1/3}a_s)+75.7\rho^{2/3}a_s^2+{\cal O}(\rho a_s^3).\label{loga}
\end{eqnarray}
The first correction agrees well with both our result and that of Ref. \cite{Arnold2001}. 
Nevertheless, two significant differences can be identified in the second term. First, our result demonstrates that this term as a polynomial and always positive, whereas in Eq. (\ref{loga}), it takes a semi-logarithmic form and is negative due to the very small gas parameter $(\rho a_s^3\ll1)$ for the dilute Bose gas. Secondly, a remarkable difference exists in the magnitude of this term. The third terms in the quadratic expression of the scattering length are identical, yet they differ substantially in magnitude.  

It is noteworthy to mention the results of Haugset {\it et al.} \cite{Haugset1998}. In their study, the CJT effective action formalism at finite temperature was employed to investigate various thermodynamic quantities, including the critical temperature. Nevertheless, the calculations were conducted within the one-loop approximation, leading to the result that the critical temperature was the same as that of the ideal Bose gas. This highlights the importance of the contribution of two-loop diagrams in determining the properties of the homogeneous weakly interacting Bose gas. 

In conclusion, we have introduced a new formalism, referred to as the CJT effective action formalism, to examine the homogeneous repulsive weakly interacting Bose gas at finite temperature. Within the framework of the IHFA, this approach satisfies three important criteria for the condensation transition of the Bose gas: (i) it exhibits a second-order phase transition; (ii) it preserves the Goldstone bosons; and (iii) in the leading-order term, the enhancement of the critical temperature is linear with respect to the scattering length. The recent advances in light shaping for optical trapping of neutral particles have led to the development of flat-bottomed optical box traps, allowing the creation of homogeneous Bose gas in the condensed phase in experiment \cite{Gaunt2013,Bause2021,Kurt2024}, we hope that our result will be verified by experimentalists.

This research is funded by Vietnam National Foundation for Science and Technology Development (NAFOSTED) under grant number 103.01-2023.12.

\bibliography{temp}

\begin{thebibliography}{40}%
\makeatletter
\providecommand \@ifxundefined [1]{%
 \@ifx{#1\undefined}
}%
\providecommand \@ifnum [1]{%
 \ifnum #1\expandafter \@firstoftwo
 \else \expandafter \@secondoftwo
 \fi
}%
\providecommand \@ifx [1]{%
 \ifx #1\expandafter \@firstoftwo
 \else \expandafter \@secondoftwo
 \fi
}%
\providecommand \natexlab [1]{#1}%
\providecommand \enquote  [1]{``#1''}%
\providecommand \bibnamefont  [1]{#1}%
\providecommand \bibfnamefont [1]{#1}%
\providecommand \citenamefont [1]{#1}%
\providecommand \href@noop [0]{\@secondoftwo}%
\providecommand \href [0]{\begingroup \@sanitize@url \@href}%
\providecommand \@href[1]{\@@startlink{#1}\@@href}%
\providecommand \@@href[1]{\endgroup#1\@@endlink}%
\providecommand \@sanitize@url [0]{\catcode `\\12\catcode `\$12\catcode
  `\&12\catcode `\#12\catcode `\^12\catcode `\_12\catcode `\%12\relax}%
\providecommand \@@startlink[1]{}%
\providecommand \@@endlink[0]{}%
\providecommand \url  [0]{\begingroup\@sanitize@url \@url }%
\providecommand \@url [1]{\endgroup\@href {#1}{\urlprefix }}%
\providecommand \urlprefix  [0]{URL }%
\providecommand \Eprint [0]{\href }%
\providecommand \doibase [0]{https://doi.org/}%
\providecommand \selectlanguage [0]{\@gobble}%
\providecommand \bibinfo  [0]{\@secondoftwo}%
\providecommand \bibfield  [0]{\@secondoftwo}%
\providecommand \translation [1]{[#1]}%
\providecommand \BibitemOpen [0]{}%
\providecommand \bibitemStop [0]{}%
\providecommand \bibitemNoStop [0]{.\EOS\space}%
\providecommand \EOS [0]{\spacefactor3000\relax}%
\providecommand \BibitemShut  [1]{\csname bibitem#1\endcsname}%
\let\auto@bib@innerbib\@empty
\bibitem [{\citenamefont {Anderson}\ \emph {et~al.}(1995)\citenamefont
  {Anderson}, \citenamefont {Ensher}, \citenamefont {Matthews}, \citenamefont
  {Wieman},\ and\ \citenamefont {Cornell}}]{Anderson1995}%
  \BibitemOpen
  \bibfield  {author} {\bibinfo {author} {\bibfnamefont {M.~H.}\ \bibnamefont
  {Anderson}}, \bibinfo {author} {\bibfnamefont {J.~R.}\ \bibnamefont
  {Ensher}}, \bibinfo {author} {\bibfnamefont {M.~R.}\ \bibnamefont
  {Matthews}}, \bibinfo {author} {\bibfnamefont {C.~E.}\ \bibnamefont
  {Wieman}},\ and\ \bibinfo {author} {\bibfnamefont {E.~A.}\ \bibnamefont
  {Cornell}},\ }\bibfield  {title} {\bibinfo {title} {Observation of
  bose-einstein condensation in a dilute atomic vapor},\ }\href
  {https://doi.org/10.1126/science.269.5221.198} {\bibfield  {journal}
  {\bibinfo  {journal} {Science}\ }\textbf {\bibinfo {volume} {269}},\ \bibinfo
  {pages} {198} (\bibinfo {year} {1995})}\BibitemShut {NoStop}%
\bibitem [{\citenamefont {Wilson}\ \emph {et~al.}(2021)\citenamefont {Wilson},
  \citenamefont {Guttridge}, \citenamefont {Segal},\ and\ \citenamefont
  {Cornish}}]{Wilson2021}%
  \BibitemOpen
  \bibfield  {author} {\bibinfo {author} {\bibfnamefont {K.~E.}\ \bibnamefont
  {Wilson}}, \bibinfo {author} {\bibfnamefont {A.}~\bibnamefont {Guttridge}},
  \bibinfo {author} {\bibfnamefont {J.}~\bibnamefont {Segal}},\ and\ \bibinfo
  {author} {\bibfnamefont {S.~L.}\ \bibnamefont {Cornish}},\ }\bibfield
  {title} {\bibinfo {title} {Quantum degenerate mixtures of cs and yb},\ }\href
  {https://doi.org/10.1103/physreva.103.033306} {\bibfield  {journal} {\bibinfo
   {journal} {Physical Review A}\ }\textbf {\bibinfo {volume} {103}},\ \bibinfo
  {pages} {033306} (\bibinfo {year} {2021})}\BibitemShut {NoStop}%
\bibitem [{\citenamefont {Burchianti}\ \emph {et~al.}(2018)\citenamefont
  {Burchianti}, \citenamefont {D’Errico}, \citenamefont {Rosi}, \citenamefont
  {Simoni}, \citenamefont {Modugno}, \citenamefont {Fort},\ and\ \citenamefont
  {Minardi}}]{Burchianti2018}%
  \BibitemOpen
  \bibfield  {author} {\bibinfo {author} {\bibfnamefont {A.}~\bibnamefont
  {Burchianti}}, \bibinfo {author} {\bibfnamefont {C.}~\bibnamefont
  {D’Errico}}, \bibinfo {author} {\bibfnamefont {S.}~\bibnamefont {Rosi}},
  \bibinfo {author} {\bibfnamefont {A.}~\bibnamefont {Simoni}}, \bibinfo
  {author} {\bibfnamefont {M.}~\bibnamefont {Modugno}}, \bibinfo {author}
  {\bibfnamefont {C.}~\bibnamefont {Fort}},\ and\ \bibinfo {author}
  {\bibfnamefont {F.}~\bibnamefont {Minardi}},\ }\bibfield  {title} {\bibinfo
  {title} {Dual-species bose-einstein condensate of k41 and rb87 in a hybrid
  trap},\ }\href {https://doi.org/10.1103/physreva.98.063616} {\bibfield
  {journal} {\bibinfo  {journal} {Physical Review A}\ }\textbf {\bibinfo
  {volume} {98}},\ \bibinfo {pages} {063616} (\bibinfo {year}
  {2018})}\BibitemShut {NoStop}%
\bibitem [{\citenamefont {Lopes}\ \emph {et~al.}(2017)\citenamefont {Lopes},
  \citenamefont {Eigen}, \citenamefont {Navon}, \citenamefont {Clément},
  \citenamefont {Smith},\ and\ \citenamefont {Hadzibabic}}]{Lopes2017}%
  \BibitemOpen
  \bibfield  {author} {\bibinfo {author} {\bibfnamefont {R.}~\bibnamefont
  {Lopes}}, \bibinfo {author} {\bibfnamefont {C.}~\bibnamefont {Eigen}},
  \bibinfo {author} {\bibfnamefont {N.}~\bibnamefont {Navon}}, \bibinfo
  {author} {\bibfnamefont {D.}~\bibnamefont {Clément}}, \bibinfo {author}
  {\bibfnamefont {R.~P.}\ \bibnamefont {Smith}},\ and\ \bibinfo {author}
  {\bibfnamefont {Z.}~\bibnamefont {Hadzibabic}},\ }\bibfield  {title}
  {\bibinfo {title} {Quantum depletion of a homogeneous bose-einstein
  condensate},\ }\href {https://doi.org/10.1103/physrevlett.119.190404}
  {\bibfield  {journal} {\bibinfo  {journal} {Physical Review Letters}\
  }\textbf {\bibinfo {volume} {119}},\ \bibinfo {pages} {190404} (\bibinfo
  {year} {2017})}\BibitemShut {NoStop}%
\bibitem [{\citenamefont {Navon}\ \emph {et~al.}(2021)\citenamefont {Navon},
  \citenamefont {Smith},\ and\ \citenamefont {Hadzibabic}}]{Navon2021}%
  \BibitemOpen
  \bibfield  {author} {\bibinfo {author} {\bibfnamefont {N.}~\bibnamefont
  {Navon}}, \bibinfo {author} {\bibfnamefont {R.~P.}\ \bibnamefont {Smith}},\
  and\ \bibinfo {author} {\bibfnamefont {Z.}~\bibnamefont {Hadzibabic}},\
  }\bibfield  {title} {\bibinfo {title} {Quantum gases in optical boxes},\
  }\href {https://doi.org/10.1038/s41567-021-01403-z} {\bibfield  {journal}
  {\bibinfo  {journal} {Nature Physics}\ }\textbf {\bibinfo {volume} {17}},\
  \bibinfo {pages} {1334} (\bibinfo {year} {2021})}\BibitemShut {NoStop}%
\bibitem [{\citenamefont {Lee}\ and\ \citenamefont {Yang}(1958)}]{Lee1958}%
  \BibitemOpen
  \bibfield  {author} {\bibinfo {author} {\bibfnamefont {T.~D.}\ \bibnamefont
  {Lee}}\ and\ \bibinfo {author} {\bibfnamefont {C.~N.}\ \bibnamefont {Yang}},\
  }\bibfield  {title} {\bibinfo {title} {Low-temperature behavior of a dilute
  bose system of hard spheres. i. equilibrium properties},\ }\href
  {https://doi.org/10.1103/physrev.112.1419} {\bibfield  {journal} {\bibinfo
  {journal} {Physical Review}\ }\textbf {\bibinfo {volume} {112}},\ \bibinfo
  {pages} {1419} (\bibinfo {year} {1958})}\BibitemShut {NoStop}%
\bibitem [{\citenamefont {Huang}(1999)}]{Huang1999}%
  \BibitemOpen
  \bibfield  {author} {\bibinfo {author} {\bibfnamefont {K.}~\bibnamefont
  {Huang}},\ }\bibfield  {title} {\bibinfo {title} {Transition temperature of a
  uniform imperfect bose gas},\ }\href
  {https://doi.org/10.1103/physrevlett.83.3770} {\bibfield  {journal} {\bibinfo
   {journal} {Physical Review Letters}\ }\textbf {\bibinfo {volume} {83}},\
  \bibinfo {pages} {3770} (\bibinfo {year} {1999})}\BibitemShut {NoStop}%
\bibitem [{\citenamefont {Mihaila}\ \emph {et~al.}(2011)\citenamefont
  {Mihaila}, \citenamefont {Cooper}, \citenamefont {Dawson}, \citenamefont
  {Chien},\ and\ \citenamefont {Timmermans}}]{Mihaila2011}%
  \BibitemOpen
  \bibfield  {author} {\bibinfo {author} {\bibfnamefont {B.}~\bibnamefont
  {Mihaila}}, \bibinfo {author} {\bibfnamefont {F.}~\bibnamefont {Cooper}},
  \bibinfo {author} {\bibfnamefont {J.~F.}\ \bibnamefont {Dawson}}, \bibinfo
  {author} {\bibfnamefont {C.-C.}\ \bibnamefont {Chien}},\ and\ \bibinfo
  {author} {\bibfnamefont {E.}~\bibnamefont {Timmermans}},\ }\bibfield  {title}
  {\bibinfo {title} {Analytical limits for cold-atom bose gases with tunable
  interactions},\ }\href {https://doi.org/10.1103/physreva.84.023603}
  {\bibfield  {journal} {\bibinfo  {journal} {Physical Review A}\ }\textbf
  {\bibinfo {volume} {84}},\ \bibinfo {pages} {023603} (\bibinfo {year}
  {2011})}\BibitemShut {NoStop}%
\bibitem [{\citenamefont {Toyoda}(1982)}]{Toyoda1982}%
  \BibitemOpen
  \bibfield  {author} {\bibinfo {author} {\bibfnamefont {T.}~\bibnamefont
  {Toyoda}},\ }\bibfield  {title} {\bibinfo {title} {A microscopic theory of
  the lambda transition},\ }\href
  {https://doi.org/10.1016/0003-4916(82)90277-9} {\bibfield  {journal}
  {\bibinfo  {journal} {Annals of Physics}\ }\textbf {\bibinfo {volume}
  {141}},\ \bibinfo {pages} {154} (\bibinfo {year} {1982})}\BibitemShut
  {NoStop}%
\bibitem [{\citenamefont {Wilkens}\ \emph {et~al.}(2000)\citenamefont
  {Wilkens}, \citenamefont {Illuminati},\ and\ \citenamefont
  {Krämer}}]{Wilkens2000}%
  \BibitemOpen
  \bibfield  {author} {\bibinfo {author} {\bibfnamefont {M.}~\bibnamefont
  {Wilkens}}, \bibinfo {author} {\bibfnamefont {F.}~\bibnamefont
  {Illuminati}},\ and\ \bibinfo {author} {\bibfnamefont {M.}~\bibnamefont
  {Krämer}},\ }\bibfield  {title} {\bibinfo {title} {Transition temperature of
  the weakly interacting bose gas: perturbative solution of the crossover
  equations in the canonical ensemble},\ }\href
  {https://doi.org/10.1088/0953-4075/33/20/10j} {\bibfield  {journal} {\bibinfo
   {journal} {Journal of Physics B: Atomic, Molecular and Optical Physics}\
  }\textbf {\bibinfo {volume} {33}},\ \bibinfo {pages} {L779} (\bibinfo {year}
  {2000})}\BibitemShut {NoStop}%
\bibitem [{\citenamefont {Bijlsma}\ and\ \citenamefont
  {Stoof}(1996)}]{Bijlsma1996}%
  \BibitemOpen
  \bibfield  {author} {\bibinfo {author} {\bibfnamefont {M.}~\bibnamefont
  {Bijlsma}}\ and\ \bibinfo {author} {\bibfnamefont {H.~T.~C.}\ \bibnamefont
  {Stoof}},\ }\bibfield  {title} {\bibinfo {title} {Renormalization group
  theory of the three-dimensional dilute bose gas},\ }\href
  {https://doi.org/10.1103/physreva.54.5085} {\bibfield  {journal} {\bibinfo
  {journal} {Physical Review A}\ }\textbf {\bibinfo {volume} {54}},\ \bibinfo
  {pages} {5085} (\bibinfo {year} {1996})}\BibitemShut {NoStop}%
\bibitem [{\citenamefont {Baym}\ \emph {et~al.}(1999)\citenamefont {Baym},
  \citenamefont {Blaizot}, \citenamefont {Holzmann}, \citenamefont {Laloë},\
  and\ \citenamefont {Vautherin}}]{Baym1999}%
  \BibitemOpen
  \bibfield  {author} {\bibinfo {author} {\bibfnamefont {G.}~\bibnamefont
  {Baym}}, \bibinfo {author} {\bibfnamefont {J.-P.}\ \bibnamefont {Blaizot}},
  \bibinfo {author} {\bibfnamefont {M.}~\bibnamefont {Holzmann}}, \bibinfo
  {author} {\bibfnamefont {F.}~\bibnamefont {Laloë}},\ and\ \bibinfo {author}
  {\bibfnamefont {D.}~\bibnamefont {Vautherin}},\ }\bibfield  {title} {\bibinfo
  {title} {The transition temperature of the dilute interacting bose gas},\
  }\href {https://doi.org/10.1103/physrevlett.83.1703} {\bibfield  {journal}
  {\bibinfo  {journal} {Physical Review Letters}\ }\textbf {\bibinfo {volume}
  {83}},\ \bibinfo {pages} {1703} (\bibinfo {year} {1999})}\BibitemShut
  {NoStop}%
\bibitem [{\citenamefont {Holzmann}\ and\ \citenamefont
  {Krauth}(1999)}]{Holzmann1999}%
  \BibitemOpen
  \bibfield  {author} {\bibinfo {author} {\bibfnamefont {M.}~\bibnamefont
  {Holzmann}}\ and\ \bibinfo {author} {\bibfnamefont {W.}~\bibnamefont
  {Krauth}},\ }\bibfield  {title} {\bibinfo {title} {Transition temperature of
  the homogeneous, weakly interacting bose gas},\ }\href
  {https://doi.org/10.1103/physrevlett.83.2687} {\bibfield  {journal} {\bibinfo
   {journal} {Physical Review Letters}\ }\textbf {\bibinfo {volume} {83}},\
  \bibinfo {pages} {2687} (\bibinfo {year} {1999})}\BibitemShut {NoStop}%
\bibitem [{\citenamefont {Arnold}\ and\ \citenamefont
  {Moore}(2001)}]{Arnold2001}%
  \BibitemOpen
  \bibfield  {author} {\bibinfo {author} {\bibfnamefont {P.}~\bibnamefont
  {Arnold}}\ and\ \bibinfo {author} {\bibfnamefont {G.}~\bibnamefont {Moore}},\
  }\bibfield  {title} {\bibinfo {title} {Bec transition temperature of a dilute
  homogeneous imperfect bose gas},\ }\href
  {https://doi.org/10.1103/physrevlett.87.120401} {\bibfield  {journal}
  {\bibinfo  {journal} {Physical Review Letters}\ }\textbf {\bibinfo {volume}
  {87}},\ \bibinfo {pages} {120401} (\bibinfo {year} {2001})}\BibitemShut
  {NoStop}%
\bibitem [{\citenamefont {Kashurnikov}\ \emph {et~al.}(2001)\citenamefont
  {Kashurnikov}, \citenamefont {Prokof’ev},\ and\ \citenamefont
  {Svistunov}}]{Kashurnikov2001}%
  \BibitemOpen
  \bibfield  {author} {\bibinfo {author} {\bibfnamefont {V.~A.}\ \bibnamefont
  {Kashurnikov}}, \bibinfo {author} {\bibfnamefont {N.~V.}\ \bibnamefont
  {Prokof’ev}},\ and\ \bibinfo {author} {\bibfnamefont {B.~V.}\ \bibnamefont
  {Svistunov}},\ }\bibfield  {title} {\bibinfo {title} {Critical temperature
  shift in weakly interacting bose gas},\ }\href
  {https://doi.org/10.1103/physrevlett.87.120402} {\bibfield  {journal}
  {\bibinfo  {journal} {Physical Review Letters}\ }\textbf {\bibinfo {volume}
  {87}},\ \bibinfo {pages} {120402} (\bibinfo {year} {2001})}\BibitemShut
  {NoStop}%
\bibitem [{\citenamefont {Baym}\ \emph {et~al.}(2001)\citenamefont {Baym},
  \citenamefont {Blaizot}, \citenamefont {Holzmann}, \citenamefont {Laloë},\
  and\ \citenamefont {Vautherin}}]{Baym2001}%
  \BibitemOpen
  \bibfield  {author} {\bibinfo {author} {\bibfnamefont {G.}~\bibnamefont
  {Baym}}, \bibinfo {author} {\bibfnamefont {J.-P.}\ \bibnamefont {Blaizot}},
  \bibinfo {author} {\bibfnamefont {M.}~\bibnamefont {Holzmann}}, \bibinfo
  {author} {\bibfnamefont {F.}~\bibnamefont {Laloë}},\ and\ \bibinfo {author}
  {\bibfnamefont {D.}~\bibnamefont {Vautherin}},\ }\bibfield  {title} {\bibinfo
  {title} {Bose-einstein transition in a dilute interacting gas},\ }\href
  {https://doi.org/10.1007/s100510170028} {\bibfield  {journal} {\bibinfo
  {journal} {The European Physical Journal B}\ }\textbf {\bibinfo {volume}
  {24}},\ \bibinfo {pages} {107} (\bibinfo {year} {2001})}\BibitemShut
  {NoStop}%
\bibitem [{\citenamefont {Holzmann}\ \emph {et~al.}(2004)\citenamefont
  {Holzmann}, \citenamefont {Fuchs}, \citenamefont {Baym}, \citenamefont
  {Blaizot},\ and\ \citenamefont {Laloë}}]{Holzmann2004}%
  \BibitemOpen
  \bibfield  {author} {\bibinfo {author} {\bibfnamefont {M.}~\bibnamefont
  {Holzmann}}, \bibinfo {author} {\bibfnamefont {J.-N.}\ \bibnamefont {Fuchs}},
  \bibinfo {author} {\bibfnamefont {G.~A.}\ \bibnamefont {Baym}}, \bibinfo
  {author} {\bibfnamefont {J.-P.}\ \bibnamefont {Blaizot}},\ and\ \bibinfo
  {author} {\bibfnamefont {F.}~\bibnamefont {Laloë}},\ }\bibfield  {title}
  {\bibinfo {title} {Bose–einstein transition temperature in a dilute
  repulsive gas},\ }\href {https://doi.org/10.1016/j.crhy.2004.01.003}
  {\bibfield  {journal} {\bibinfo  {journal} {Comptes Rendus. Physique}\
  }\textbf {\bibinfo {volume} {5}},\ \bibinfo {pages} {21} (\bibinfo {year}
  {2004})}\BibitemShut {NoStop}%
\bibitem [{\citenamefont {Holzmann}\ \emph {et~al.}(1999)\citenamefont
  {Holzmann}, \citenamefont {Grüter},\ and\ \citenamefont
  {Laloë}}]{Holzmann1999a}%
  \BibitemOpen
  \bibfield  {author} {\bibinfo {author} {\bibfnamefont {M.}~\bibnamefont
  {Holzmann}}, \bibinfo {author} {\bibfnamefont {P.}~\bibnamefont {Grüter}},\
  and\ \bibinfo {author} {\bibfnamefont {F.}~\bibnamefont {Laloë}},\
  }\bibfield  {title} {\bibinfo {title} {Bose-einstein condensation in
  interacting gases},\ }\href {https://doi.org/10.1007/s100510050905}
  {\bibfield  {journal} {\bibinfo  {journal} {The European Physical Journal B}\
  }\textbf {\bibinfo {volume} {10}},\ \bibinfo {pages} {739} (\bibinfo {year}
  {1999})}\BibitemShut {NoStop}%
\bibitem [{\citenamefont {Grüter}\ \emph {et~al.}(1997)\citenamefont
  {Grüter}, \citenamefont {Ceperley},\ and\ \citenamefont
  {Laloë}}]{Grueter1997}%
  \BibitemOpen
  \bibfield  {author} {\bibinfo {author} {\bibfnamefont {P.}~\bibnamefont
  {Grüter}}, \bibinfo {author} {\bibfnamefont {D.}~\bibnamefont {Ceperley}},\
  and\ \bibinfo {author} {\bibfnamefont {F.}~\bibnamefont {Laloë}},\
  }\bibfield  {title} {\bibinfo {title} {Critical temperature of bose-einstein
  condensation of hard-sphere gases},\ }\href
  {https://doi.org/10.1103/physrevlett.79.3549} {\bibfield  {journal} {\bibinfo
   {journal} {Physical Review Letters}\ }\textbf {\bibinfo {volume} {79}},\
  \bibinfo {pages} {3549} (\bibinfo {year} {1997})}\BibitemShut {NoStop}%
\bibitem [{\citenamefont {de~Souza~Cruz}\ \emph {et~al.}(2001)\citenamefont
  {de~Souza~Cruz}, \citenamefont {Pinto},\ and\ \citenamefont
  {Ramos}}]{SouzaCruz2001}%
  \BibitemOpen
  \bibfield  {author} {\bibinfo {author} {\bibfnamefont {F.~F.}\ \bibnamefont
  {de~Souza~Cruz}}, \bibinfo {author} {\bibfnamefont {M.~B.}\ \bibnamefont
  {Pinto}},\ and\ \bibinfo {author} {\bibfnamefont {R.~O.}\ \bibnamefont
  {Ramos}},\ }\bibfield  {title} {\bibinfo {title} {Transition temperature for
  weakly interacting homogeneous bose gases},\ }\href
  {https://doi.org/10.1103/physrevb.64.014515} {\bibfield  {journal} {\bibinfo
  {journal} {Physical Review B}\ }\textbf {\bibinfo {volume} {64}},\ \bibinfo
  {pages} {014515} (\bibinfo {year} {2001})}\BibitemShut {NoStop}%
\bibitem [{\citenamefont {Cornwall}\ \emph {et~al.}(1974)\citenamefont
  {Cornwall}, \citenamefont {Jackiw},\ and\ \citenamefont
  {Tomboulis}}]{Cornwall1974}%
  \BibitemOpen
  \bibfield  {author} {\bibinfo {author} {\bibfnamefont {J.~M.}\ \bibnamefont
  {Cornwall}}, \bibinfo {author} {\bibfnamefont {R.}~\bibnamefont {Jackiw}},\
  and\ \bibinfo {author} {\bibfnamefont {E.}~\bibnamefont {Tomboulis}},\
  }\bibfield  {title} {\bibinfo {title} {Effective action for composite
  operators},\ }\href {https://doi.org/10.1103/physrevd.10.2428} {\bibfield
  {journal} {\bibinfo  {journal} {Physical Review D}\ }\textbf {\bibinfo
  {volume} {10}},\ \bibinfo {pages} {2428} (\bibinfo {year}
  {1974})}\BibitemShut {NoStop}%
\bibitem [{\citenamefont {Thu}\ and\ \citenamefont {Pham}(2024)}]{Thu2024}%
  \BibitemOpen
  \bibfield  {author} {\bibinfo {author} {\bibfnamefont {N.~V.}\ \bibnamefont
  {Thu}}\ and\ \bibinfo {author} {\bibfnamefont {D.~T.}\ \bibnamefont {Pham}},\
  }\bibfield  {title} {\bibinfo {title} {Effect of nonzero temperature to
  non-condensed fraction of a homogeneous dilute weakly interacting bose gas},\
  }\href {https://doi.org/10.1016/j.physleta.2024.129787} {\bibfield  {journal}
  {\bibinfo  {journal} {Physics Letters A}\ }\textbf {\bibinfo {volume}
  {523}},\ \bibinfo {pages} {129787} (\bibinfo {year} {2024})}\BibitemShut
  {NoStop}%
\bibitem [{\citenamefont {Pethick}\ and\ \citenamefont
  {Smith}(2008)}]{Pethick2008}%
  \BibitemOpen
  \bibfield  {author} {\bibinfo {author} {\bibfnamefont {C.~J.}\ \bibnamefont
  {Pethick}}\ and\ \bibinfo {author} {\bibfnamefont {H.}~\bibnamefont
  {Smith}},\ }\href@noop {} {\emph {\bibinfo {title} {Bose-Einstein
  Condensation in Dilute Gases}}}\ (\bibinfo  {publisher} {Cambridge University
  Press},\ \bibinfo {year} {2008})\BibitemShut {NoStop}%
\bibitem [{\citenamefont {Pitaevskii}\ and\ \citenamefont
  {Stringari}(2003)}]{Pitaevskii2003}%
  \BibitemOpen
  \bibfield  {author} {\bibinfo {author} {\bibfnamefont {L.~P.}\ \bibnamefont
  {Pitaevskii}}\ and\ \bibinfo {author} {\bibfnamefont {S.}~\bibnamefont
  {Stringari}},\ }\href@noop {} {\emph {\bibinfo {title} {Bose-Einstein
  Condensation}}}\ (\bibinfo  {publisher} {Oxford University Press},\ \bibinfo
  {year} {2003})\BibitemShut {NoStop}%
\bibitem [{\citenamefont {Van~Thu}\ and\ \citenamefont
  {Berx}(2022)}]{VanThu2022}%
  \BibitemOpen
  \bibfield  {author} {\bibinfo {author} {\bibfnamefont {N.}~\bibnamefont
  {Van~Thu}}\ and\ \bibinfo {author} {\bibfnamefont {J.}~\bibnamefont {Berx}},\
  }\bibfield  {title} {\bibinfo {title} {The condensed fraction of a
  homogeneous dilute bose gas within the improved hartree–fock
  approximation},\ }\bibfield  {journal} {\bibinfo  {journal} {Journal of
  Statistical Physics}\ }\textbf {\bibinfo {volume} {188}},\ \href
  {https://doi.org/10.1007/s10955-022-02944-0} {10.1007/s10955-022-02944-0}
  (\bibinfo {year} {2022})\BibitemShut {NoStop}%
\bibitem [{\citenamefont {Anderson}(2018)}]{Anderson2018}%
  \BibitemOpen
  \bibfield  {author} {\bibinfo {author} {\bibfnamefont {P.~W.}\ \bibnamefont
  {Anderson}},\ }\href {https://doi.org/10.4324/9780429494116} {\emph {\bibinfo
  {title} {Basic Notions of Condensed Matter Physics}}},\ edited by\ \bibinfo
  {editor} {\bibfnamefont {P.~W.}\ \bibnamefont {Anderson}}\ (\bibinfo
  {publisher} {CRC Press},\ \bibinfo {year} {2018})\BibitemShut {NoStop}%
\bibitem [{\citenamefont {Goldstone}(1961)}]{Goldstone1961}%
  \BibitemOpen
  \bibfield  {author} {\bibinfo {author} {\bibfnamefont {J.}~\bibnamefont
  {Goldstone}},\ }\bibfield  {title} {\bibinfo {title} {Field theories with
  "superconductor" solutions},\ }\href {https://doi.org/10.1007/bf02812722}
  {\bibfield  {journal} {\bibinfo  {journal} {Il Nuovo Cimento}\ }\textbf
  {\bibinfo {volume} {19}},\ \bibinfo {pages} {154} (\bibinfo {year}
  {1961})}\BibitemShut {NoStop}%
\bibitem [{\citenamefont {Hugenholtz}\ and\ \citenamefont
  {Pines}(1959)}]{Hugenholtz1959}%
  \BibitemOpen
  \bibfield  {author} {\bibinfo {author} {\bibfnamefont {N.~M.}\ \bibnamefont
  {Hugenholtz}}\ and\ \bibinfo {author} {\bibfnamefont {D.}~\bibnamefont
  {Pines}},\ }\bibfield  {title} {\bibinfo {title} {Ground-state energy and
  excitation spectrum of a system of interacting bosons},\ }\href
  {https://doi.org/10.1103/physrev.116.489} {\bibfield  {journal} {\bibinfo
  {journal} {Physical Review}\ }\textbf {\bibinfo {volume} {116}},\ \bibinfo
  {pages} {489} (\bibinfo {year} {1959})}\BibitemShut {NoStop}%
\bibitem [{\citenamefont {Hohenberg}\ and\ \citenamefont
  {Martin}(1965)}]{Hohenberg1965}%
  \BibitemOpen
  \bibfield  {author} {\bibinfo {author} {\bibfnamefont {P.}~\bibnamefont
  {Hohenberg}}\ and\ \bibinfo {author} {\bibfnamefont {P.}~\bibnamefont
  {Martin}},\ }\bibfield  {title} {\bibinfo {title} {Microscopic theory of
  superfluid helium},\ }\href {https://doi.org/10.1016/0003-4916(65)90280-0}
  {\bibfield  {journal} {\bibinfo  {journal} {Annals of Physics}\ }\textbf
  {\bibinfo {volume} {34}},\ \bibinfo {pages} {291} (\bibinfo {year}
  {1965})}\BibitemShut {NoStop}%
\bibitem [{\citenamefont {Schmitt}(2010)}]{Schmitt2010}%
  \BibitemOpen
  \bibfield  {author} {\bibinfo {author} {\bibfnamefont {A.}~\bibnamefont
  {Schmitt}},\ }\href@noop {} {\emph {\bibinfo {title} {Dense matter in compact
  stars}}}\ (\bibinfo  {publisher} {Springer-Verlag Berlin Heidelberg},\
  \bibinfo {year} {2010})\BibitemShut {NoStop}%
\bibitem [{\citenamefont {Andersen}(2004)}]{Andersen2004}%
  \BibitemOpen
  \bibfield  {author} {\bibinfo {author} {\bibfnamefont {J.}~\bibnamefont
  {Andersen}},\ }\bibfield  {title} {\bibinfo {title} {Theory of the weakly
  interacting bose gas},\ }\href {https://doi.org/10.1103/revmodphys.76.599}
  {\bibfield  {journal} {\bibinfo  {journal} {Reviews of Modern Physics}\
  }\textbf {\bibinfo {volume} {76}},\ \bibinfo {pages} {599} (\bibinfo {year}
  {2004})}\BibitemShut {NoStop}%
\bibitem [{\citenamefont {Olver}\ \emph {et~al.}(2010)\citenamefont {Olver},
  \citenamefont {Lozier}, \citenamefont {Boisvert},\ and\ \citenamefont
  {Clark}}]{Olver2010}%
  \BibitemOpen
  \bibinfo {editor} {\bibfnamefont {F.~W.~J.}\ \bibnamefont {Olver}}, \bibinfo
  {editor} {\bibfnamefont {D.~W.}\ \bibnamefont {Lozier}}, \bibinfo {editor}
  {\bibfnamefont {R.~F.}\ \bibnamefont {Boisvert}},\ and\ \bibinfo {editor}
  {\bibfnamefont {C.~W.}\ \bibnamefont {Clark}},\ eds.,\ \href@noop {} {\emph
  {\bibinfo {title} {NIST handbook of mathematical functions}}}\ (\bibinfo
  {publisher} {Cambridge University Press},\ \bibinfo {address} {Cambridge},\
  \bibinfo {year} {2010})\ \bibinfo {note} {neubearbeitung von: Handbook of
  mathematical functions with formulas, graphs, and mathematical tables / M.
  Abramowitz and I.A. Stegun, editors (1964)}\BibitemShut {NoStop}%
\bibitem [{\citenamefont {Shi}(1998)}]{Shi1998}%
  \BibitemOpen
  \bibfield  {author} {\bibinfo {author} {\bibfnamefont {H.}~\bibnamefont
  {Shi}},\ }\bibfield  {title} {\bibinfo {title} {Finite-temperature
  excitations in a dilute bose-condensed gas},\ }\href
  {https://doi.org/10.1016/s0370-1573(98)00015-5} {\bibfield  {journal}
  {\bibinfo  {journal} {Physics Reports}\ }\textbf {\bibinfo {volume} {304}},\
  \bibinfo {pages} {1} (\bibinfo {year} {1998})}\BibitemShut {NoStop}%
\bibitem [{\citenamefont {Kastening}(2004)}]{Kastening2004}%
  \BibitemOpen
  \bibfield  {author} {\bibinfo {author} {\bibfnamefont {B.}~\bibnamefont
  {Kastening}},\ }\bibfield  {title} {\bibinfo {title} {Bose-einstein
  condensation temperature of a homogenous weakly interacting bose gas in
  variational perturbation theory through seven loops},\ }\href
  {https://doi.org/10.1103/physreva.69.043613} {\bibfield  {journal} {\bibinfo
  {journal} {Physical Review A}\ }\textbf {\bibinfo {volume} {69}},\ \bibinfo
  {pages} {043613} (\bibinfo {year} {2004})}\BibitemShut {NoStop}%
\bibitem [{\citenamefont {Holzmann}\ \emph {et~al.}(2001)\citenamefont
  {Holzmann}, \citenamefont {Baym}, \citenamefont {Blaizot},\ and\
  \citenamefont {Laloë}}]{Holzmann2001}%
  \BibitemOpen
  \bibfield  {author} {\bibinfo {author} {\bibfnamefont {M.}~\bibnamefont
  {Holzmann}}, \bibinfo {author} {\bibfnamefont {G.}~\bibnamefont {Baym}},
  \bibinfo {author} {\bibfnamefont {J.-P.}\ \bibnamefont {Blaizot}},\ and\
  \bibinfo {author} {\bibfnamefont {F.}~\bibnamefont {Laloë}},\ }\bibfield
  {title} {\bibinfo {title} {Nonanalytic dependence of the transition
  temperature of the homogeneous dilute bose gas on scattering length},\ }\href
  {https://doi.org/10.1103/physrevlett.87.120403} {\bibfield  {journal}
  {\bibinfo  {journal} {Physical Review Letters}\ }\textbf {\bibinfo {volume}
  {87}},\ \bibinfo {pages} {120403} (\bibinfo {year} {2001})}\BibitemShut
  {NoStop}%
\bibitem [{\citenamefont {Arnold}\ \emph {et~al.}(2001)\citenamefont {Arnold},
  \citenamefont {Moore},\ and\ \citenamefont {Tomášik}}]{Arnold2001a}%
  \BibitemOpen
  \bibfield  {author} {\bibinfo {author} {\bibfnamefont {P.}~\bibnamefont
  {Arnold}}, \bibinfo {author} {\bibfnamefont {G.}~\bibnamefont {Moore}},\ and\
  \bibinfo {author} {\bibfnamefont {B.}~\bibnamefont {Tomášik}},\ }\bibfield
  {title} {\bibinfo {title} {Tcfor homogeneous dilute bose gases: A
  second-order result},\ }\href {https://doi.org/10.1103/physreva.65.013606}
  {\bibfield  {journal} {\bibinfo  {journal} {Physical Review A}\ }\textbf
  {\bibinfo {volume} {65}},\ \bibinfo {pages} {013606} (\bibinfo {year}
  {2001})}\BibitemShut {NoStop}%
\bibitem [{\citenamefont {Haugset}\ \emph {et~al.}(1998)\citenamefont
  {Haugset}, \citenamefont {Haugerud},\ and\ \citenamefont
  {Ravndal}}]{Haugset1998}%
  \BibitemOpen
  \bibfield  {author} {\bibinfo {author} {\bibfnamefont {T.}~\bibnamefont
  {Haugset}}, \bibinfo {author} {\bibfnamefont {H.}~\bibnamefont {Haugerud}},\
  and\ \bibinfo {author} {\bibfnamefont {F.}~\bibnamefont {Ravndal}},\
  }\bibfield  {title} {\bibinfo {title} {Thermodynamics of a weakly interacting
  bose–einstein gas},\ }\href {https://doi.org/10.1006/aphy.1998.5795}
  {\bibfield  {journal} {\bibinfo  {journal} {Annals of Physics}\ }\textbf
  {\bibinfo {volume} {266}},\ \bibinfo {pages} {27} (\bibinfo {year}
  {1998})}\BibitemShut {NoStop}%
\bibitem [{\citenamefont {Gaunt}\ \emph {et~al.}(2013)\citenamefont {Gaunt},
  \citenamefont {Schmidutz}, \citenamefont {Gotlibovych}, \citenamefont
  {Smith},\ and\ \citenamefont {Hadzibabic}}]{Gaunt2013}%
  \BibitemOpen
  \bibfield  {author} {\bibinfo {author} {\bibfnamefont {A.~L.}\ \bibnamefont
  {Gaunt}}, \bibinfo {author} {\bibfnamefont {T.~F.}\ \bibnamefont
  {Schmidutz}}, \bibinfo {author} {\bibfnamefont {I.}~\bibnamefont
  {Gotlibovych}}, \bibinfo {author} {\bibfnamefont {R.~P.}\ \bibnamefont
  {Smith}},\ and\ \bibinfo {author} {\bibfnamefont {Z.}~\bibnamefont
  {Hadzibabic}},\ }\bibfield  {title} {\bibinfo {title} {Bose-einstein
  condensation of atoms in a uniform potential},\ }\href
  {https://doi.org/10.1103/physrevlett.110.200406} {\bibfield  {journal}
  {\bibinfo  {journal} {Physical Review Letters}\ }\textbf {\bibinfo {volume}
  {110}},\ \bibinfo {pages} {200406} (\bibinfo {year} {2013})}\BibitemShut
  {NoStop}%
\bibitem [{\citenamefont {Bause}\ \emph {et~al.}(2021)\citenamefont {Bause},
  \citenamefont {Schindewolf}, \citenamefont {Tao}, \citenamefont {Duda},
  \citenamefont {Chen}, \citenamefont {Quéméner}, \citenamefont {Karman},
  \citenamefont {Christianen}, \citenamefont {Bloch},\ and\ \citenamefont
  {Luo}}]{Bause2021}%
  \BibitemOpen
  \bibfield  {author} {\bibinfo {author} {\bibfnamefont {R.}~\bibnamefont
  {Bause}}, \bibinfo {author} {\bibfnamefont {A.}~\bibnamefont {Schindewolf}},
  \bibinfo {author} {\bibfnamefont {R.}~\bibnamefont {Tao}}, \bibinfo {author}
  {\bibfnamefont {M.}~\bibnamefont {Duda}}, \bibinfo {author} {\bibfnamefont
  {X.-Y.}\ \bibnamefont {Chen}}, \bibinfo {author} {\bibfnamefont
  {G.}~\bibnamefont {Quéméner}}, \bibinfo {author} {\bibfnamefont
  {T.}~\bibnamefont {Karman}}, \bibinfo {author} {\bibfnamefont
  {A.}~\bibnamefont {Christianen}}, \bibinfo {author} {\bibfnamefont
  {I.}~\bibnamefont {Bloch}},\ and\ \bibinfo {author} {\bibfnamefont {X.-Y.}\
  \bibnamefont {Luo}},\ }\bibfield  {title} {\bibinfo {title} {Collisions of
  ultracold molecules in bright and dark optical dipole traps},\ }\href
  {https://doi.org/10.1103/physrevresearch.3.033013} {\bibfield  {journal}
  {\bibinfo  {journal} {Physical Review Research}\ }\textbf {\bibinfo {volume}
  {3}},\ \bibinfo {pages} {033013} (\bibinfo {year} {2021})}\BibitemShut
  {NoStop}%
\bibitem [{\citenamefont {Kurt}\ \emph {et~al.}(2024)\citenamefont {Kurt},
  \citenamefont {Sisman},\ and\ \citenamefont {Aydin}}]{Kurt2024}%
  \BibitemOpen
  \bibfield  {author} {\bibinfo {author} {\bibfnamefont {C.}~\bibnamefont
  {Kurt}}, \bibinfo {author} {\bibfnamefont {A.}~\bibnamefont {Sisman}},\ and\
  \bibinfo {author} {\bibfnamefont {A.}~\bibnamefont {Aydin}},\ }\href
  {https://doi.org/10.48550/ARXIV.2408.12698} {\bibinfo {title} {Shape-induced
  bose-einstein condensation}} (\bibinfo {year} {2024})\BibitemShut {NoStop}%
\end{thebibliography}%

\end{document}